\documentclass[%
 reprint,
 superscriptaddress,
 amsmath,amssymb,
 aps,
 prl,
]{revtex4-1}

\usepackage{graphicx}
\usepackage{dcolumn}
\usepackage{xcolor}
\usepackage{bm}

\usepackage[colorlinks=true,linkcolor=blue, citecolor=blue]{hyperref}%

\begin{document}

\title{Hidden Interplay of Current-Induced Spin and Orbital Torques in Bulk Fe$_3$GeTe$_2$}

\author{Tom G. Saunderson}
\email{tsaunder@uni-mainz.de}
\affiliation{Institute of Physics, Johannes Gutenberg-University, 55099 Mainz, Germany}
\affiliation{Peter Gr\"unberg Institut and Institute for Advanced Simulation, Forschungszentrum J\"ulich and JARA, 52425 J\"ulich, Germany} 
\author{Dongwook Go}%
\affiliation{Institute of Physics, Johannes Gutenberg-University, 55099 Mainz, Germany}
\affiliation{Peter Gr\"unberg Institut and Institute for Advanced Simulation, Forschungszentrum J\"ulich and JARA, 52425 J\"ulich, Germany} 
\author{Stefan Bl\"ugel}%
\affiliation{Peter Gr\"unberg Institut and Institute for Advanced Simulation, Forschungszentrum J\"ulich and JARA, 52425 J\"ulich, Germany} 
\author{Mathias Kl\"aui}
\affiliation{Institute of Physics, Johannes Gutenberg-University, 55099 Mainz, Germany}
\affiliation{Centre for Quantum Spintronics, Department of Physics, Norwegian University of Science and Technology, 7491 Trondheim, Norway}
\author{Yuriy Mokrousov}%
\affiliation{Institute of Physics, Johannes Gutenberg-University, 55099 Mainz, Germany}
\affiliation{Peter Gr\"unberg Institut and Institute for Advanced Simulation, Forschungszentrum J\"ulich and JARA, 52425 J\"ulich, Germany}

\date{\today}

\begin{abstract}

Low crystal symmetry of magnetic van der Waals materials naturally promotes spin-orbital complexity unachievable in common magnetic materials used for spin-orbit torque switching.  Here, using first-principles methods, we demonstrate that an interplay of spin and orbital degrees of freedom has a profound impact on spin-orbit torques in the prototypical van der Waals ferromagnet: Fe$_3$GeTe$_2$. 
While we show that bulk Fe$_3$GeTe$_2$ hosts strong ``hidden" current-induced torques harvested by each of its layers, we uncover that their origin alternates between the conventional spin flux torque and  the so-called orbital torque as the magnetization direction is varied.
A drastic difference in the behavior of the two types of torques results in a non-trivial evolution of switching properties with doping. Our findings promote the design of non-equilibrium orbital properties as the guiding mechanism for crafting the properties of spin-orbit torques in layered van der Waals materials.


\end{abstract}

\maketitle

The discovery of two-dimensional (2D) van der Waals (vdW) ferromagnets Cr$_2$Ge$_2$Te$_6$ \cite{Gong2017} and CrI$_3$ \cite{Huang2017} has been long-awaited since the works on how spatial dimensionality affects criticality \cite{Griffiths1964,Mermin1966}. Now, the potential  applications of these discoveries seem vast within the field of spintronics \cite{Hu2020a}. A 2D layered material could enable the \textcolor{black}{more} efficient design of novel spintronic devices \textcolor{black}{than their metallic bilayer counterparts with similar symmetries \cite{Liu2021}}. The recent discovery of Fe$_3$GeTe$_2$ (FGT) \cite{Deiseroth2006} stands as a significant milestone since its Curie temperature could be raised to room temperature with ionic liquid gating \cite{Deng2018}. FGT has since been researched extensively, showing exciting characteristics such as nodal line semimetallicity \cite{Kim2018b}, skyrmionic spin textures~\cite{Ding2020, Park2021} and controllable spin currents \cite{Zhou2021}. In particular, experiments have demonstrated magnetization switching in Pt/FGT heterostructures via current-induced spin-orbit torque \cite{Alghamdi2019,Zhang2021a}.

\begin{figure}
\includegraphics*[width=0.9\linewidth,clip]{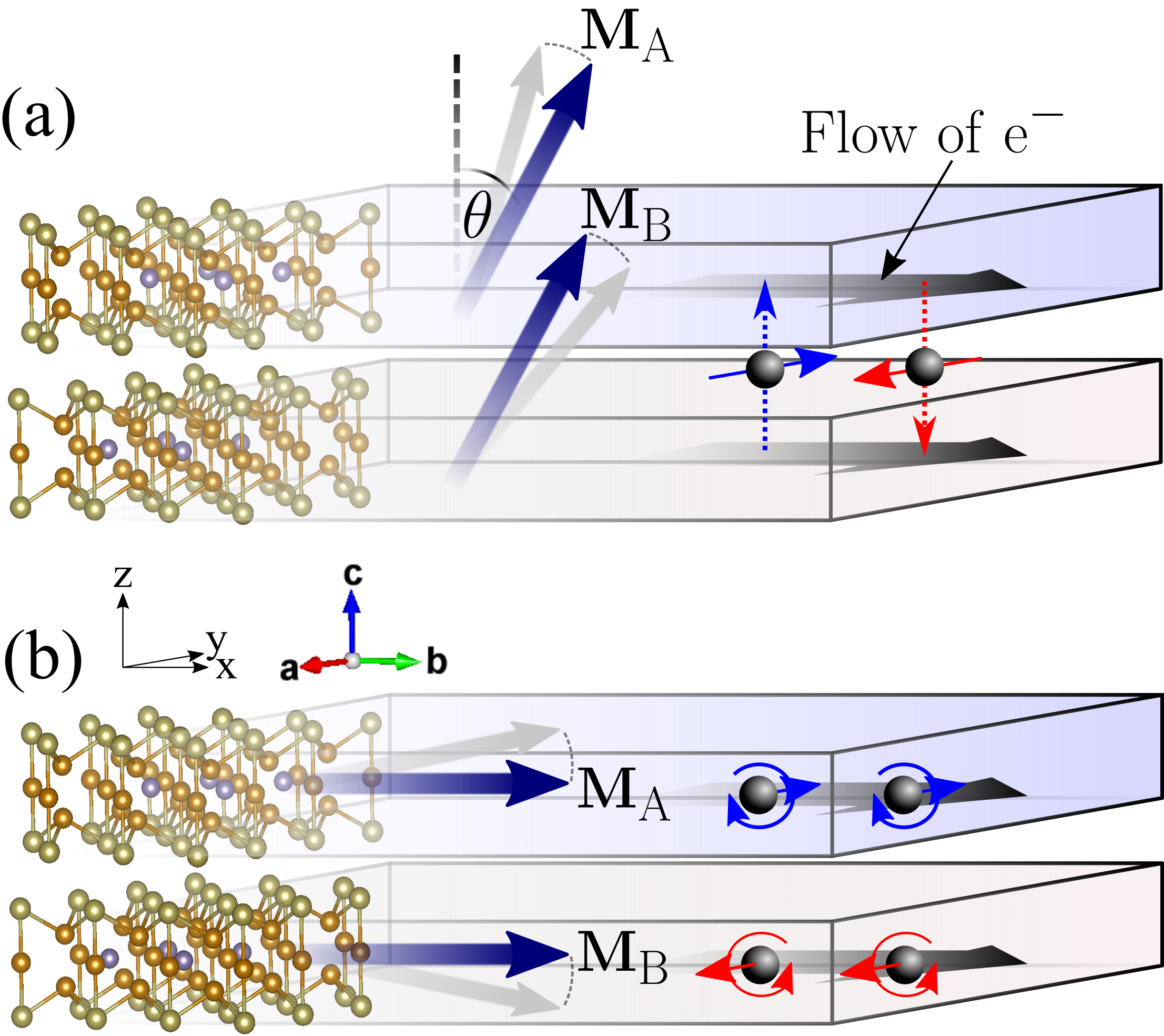}
\caption{A schematic illustration for current-induced torques in Fe$_3$GeTe$_2$ (FGT) with (a) an out-of-plane and (b) in-plane magnetization. In both instances the current direction
is along the $x$-axis. The crystal structure of Fe$_3$GeTe$_2$, consisting of two individual layers, A and B, is displayed on the left. 
The direction of the magnetization at angle $\theta$ with the $z$-axis is marked with $\mathbf{M}_\mathrm{A}$ and $\mathbf{M}_\mathrm{B}$ for each layer.
The current-induced torques on each layer are opposite, so that the overall torque is vanishing. While for out-of-plane FGT, the magnetic torque is mainly driven by the flow of the spin angular momentum between layers [marked with arrows in (a)], the magnetic torque is dominated by prominant orbital accumulation for in-plane FGT [marked with arrows in (b)].
} 
\label{fig:OutofPlaneInPlane}
\end{figure}

We note that the magnetization switching demonstrated in Refs.~\cite{Alghamdi2019, Zhang2021a} utilized the spin-Hall effect in Pt, where the physical principle is analogous to conventional SOT devices with transition-metal bilayers. A unique role played by the low crystal symmetry of FGT was pointed out by Johansen {\it et al.} \cite{Johansen2019}, which allows for the generation of a current-induced torque without the need for a heavy metal interface. Experiments have now established the existence of such an unconventional torque \cite{Zhang2021, Martin2021}. However, these experiments were performed on bulk-like thick samples $\gtrsim 20\ \mathrm{nm}$ of FGT. A key aspect of bulk FGT is the global inversion symmetry that prevents net torque response. However two vdW layers in the unit cell of bulk FGT, referred to as A and B layers, are inversion partners with respect to each other, see Fig.~\ref{fig:OutofPlaneInPlane}. Consequently despite total torque vanishing, individual layers host local ``staggered'' torques. We note that a similar situation is encountered in other centrosymmetric systems, for which the concept of the ``hidden" Rashba effect has been developed ~\cite{Yuan2019,Atkinson2020,Lee2020}. 

Although the symmetry dictates that current-induced torque in bulk FGT is similar to multiple copies of the result for monolayer FGT \cite{Zhang2021}, the microscopic mechanism in bulk FGT is expected to be different from that of the monolayer. For instance, while current-induced torque in monolayer FGT will consist of both spin and orbital accumulations, in bulk FGT, spin and orbital currents exchanged between layers are also expected to contribute. In particular, recent theories predict that the orbital current can be generated much more efficiently than the spin current, e.g. via orbital Hall effect, not only in transition metals \cite{Kontani2008, Kontani2009a,Go2018a,Jo2018, Salemi2022, Salemi2022b} but also in 2D materials \cite{Canonico2020a, Canonico2020b, Bhowal20020b, Bhowal20020b, Cysne2021, Bhowal2021}. Moreover, the orbital current can exert torque on magnetic moments by transferring orbital angular momentum \cite{Go2020a, Go2020}. Known as the orbital torque, it provides a promising alternative route to control the magnetization through the orbital degree of freedom, and is considered a fresh perspective \cite{EPL_review_orbitronics}. So far, the orbital torque has been experimentally observed mostly in transition-metal-based magnetic heterostructures \cite{Ding2020a, Kim2021, Lee2021a, Lee2021b, Ding2022, Liao2022, Hu2022, Hayashi2022, Santos2022}. However, owing to a highly anisotropic crystal field potential, strong spin-orbit coupling (SOC), and orbital complexity of the electronic structure, bulk FGT is expected to exhibit rich entangled dynamics between spin and orbital degrees of freedom. This motivates our search for the role of orbital excitations and its interaction with magnetic moments in bulk FGT, which will enhance the currently underdeveloped understanding of orbital physics in vdW materials. 

In this work, by employing first principles methods, we unveil interplay of the spin and orbital angular momentum in bulk FGT, which is ``hidden'' by the global inversion symmetry. When the magnetization is pointing out of the vdW plane, we find spin flux exchanged between vdW layers is the major contribution to the torque on the magnetization, Fig.~\ref{fig:OutofPlaneInPlane}(a). In contrast, as the magnetization direction gets closer to the vdW plane, current-induced orbital accumulation becomes strongly pronounced, which leads to a large orbital torque response, Fig.~\ref{fig:OutofPlaneInPlane}(b). This is the first example of an intrinsic crossover between spin and orbital torques, making bulk FGT an ideal candidate for studying non-equilibrium orbital excitations.

\begin{figure}[t!]
\includegraphics*[width=1\linewidth,clip]{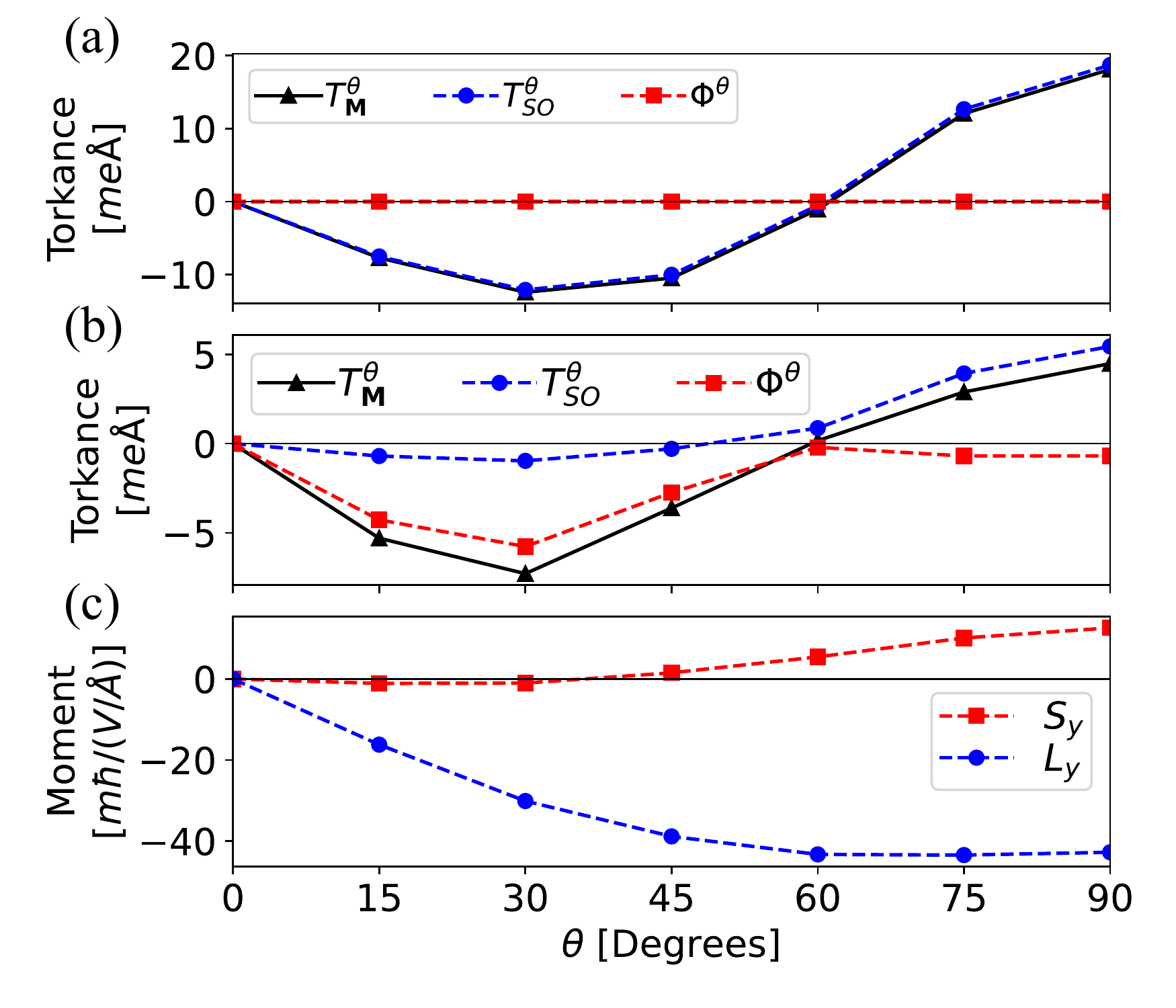}
\caption{Angular dependence of current-induced torque and the spin and orbital contributions in FGT. 
(a) The  $\theta$-component of the anti-damping current-induced torque on the magnetization ($T^\theta_\mathbf{M}$, black triangles), the spin contribution ($\varPhi^{\theta}$, red dashed square), and the orbital contribution ($T^\theta_{SO}$, blue dashed circle) for single-layer FGT as a function of $\theta$. 
(b) Same as in (a) for the torkance projected onto the A-layer in bulk FGT.
(c) The $y$-component of the current-induced orbital ($L_y$, blue dashed circle) and spin ($S_y$, red dashed square) moment summed over Fe atoms in A-layer of bulk FGT as a function of the angle $\theta$. 
}
\label{fig:AngularDepSOT}
\end{figure}

\begin{figure*}[t!]
\includegraphics*[width=1\linewidth,clip]{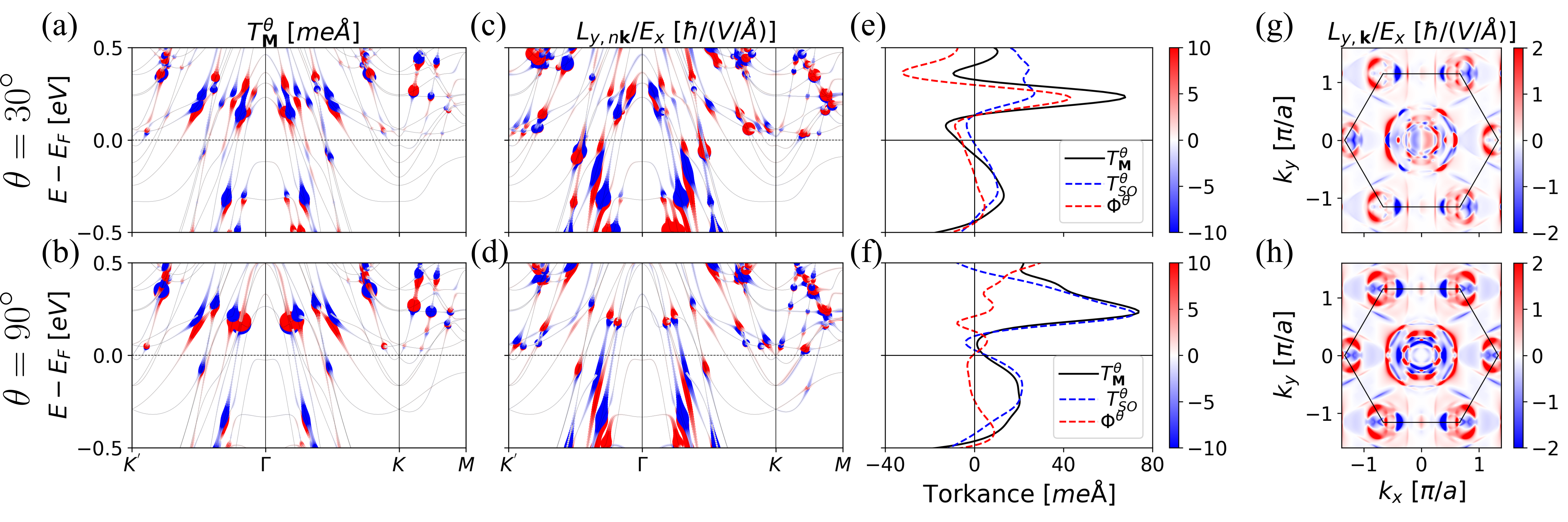}
\caption{Properties of the current-induced torque and orbital angular momentum of the A-layer of Fe$_3$GeTe$_2$ for the  magnetization angle of $\theta=30^\circ$ [panels (a, c, e, g)] and $\theta=90^\circ$ [panels  (b,d,f,h)]. 
In panels (a,b) the plotted band structure is superimposed with  the coloured circles whose color and size represents the  expectation value of the total torkance $T^{\theta}_{\mathbf{M},n\mathbf{k}}$. Similarly, in (c,d) it is the $y$-component of the current-induced orbital angular momentum  $L_{y,n\mathbf{k}}$ which is represented with the circles. Additionally, the $\mathbf{k}$-space distribution of $L_{y,\mathbf{k}}$ for the two angles is shown in (g,h).  Panels (e,f) display the total integrated  ($T^\theta$, solid line), flux ($\varPhi^\theta$, red dashed line) and orbital ($T_{\rm SO}^\theta$, blue dashed line) torkance as a function of band filling for $\theta=30^\circ$ (e) and $\theta=90^\circ$ (f), where $E=0$ represents the true Fermi energy of FGT.
The prominent current-induced orbital angular momentum and orbital torque for the in-plane magnetization is evident.
} 
\label{fig:TorquesOnBandsAndEDepSOT}
\end{figure*}


To describe the current-induced exchange of spin and orbital angular momentum between A and B layers, which results in torques on the magnetization, we adopt the spin continuity equation~\cite{Go2020}: 
\begin{eqnarray}
\label{eq:continuity}
\frac{d\mathbf{S}_\mathrm{A}}{dt}
=
\boldsymbol{\varPhi}_\mathrm{A}
+
\boldsymbol{\mathcal{T}}_\mathrm{SO,A}
- \boldsymbol{\mathcal{T}}_{\mathbf{M},\mathrm{A}},
\end{eqnarray}
where the subscript stands for the vdW layer A. Here, $\boldsymbol{\varPhi}_\mathrm{A}$ is the spin flux from layer B into layer A, $\boldsymbol{\mathcal{T}}_\mathrm{SO,A} \sim \mathbf{L}_{\mathrm{A}} \times \mathbf{S}_{\mathrm{A}}$ describes mutual precession between the spin ($\mathbf{S}_{\mathrm{A}}$) and orbital ($\mathbf{L}_{\mathrm{A}}$) angular momenta at layer A via SOC, and $\boldsymbol{\mathcal{T}}_{\mathbf{M},\mathrm{A}} \sim \hat{\mathbf{M}}_{\rm A}\times \mathbf{S}_{\mathrm{A}}$ is the torque that the spin $\mathbf{S}_{\mathrm{A}}$ exerts on A-layer magnetization $\mathbf{M}_{\rm A}$ through the exchange interaction. 
We remark that the spin flux and local torques on layers A and B have the same magnitude but are opposite in sign  by inversion symmetry. Thus, we analyze the dynamics of the A-layer only and remove the subscript A in the following.
In a steady state, when ${d\mathbf{S}_\mathrm{A}}/{dt}
= 0$, current-induced torque on the magnetization is given by
two contributions: the spin flux $ \boldsymbol{\varPhi}$, and the angular momentum transfer from orbital to spin $\boldsymbol{\mathcal{T}}_{\mathrm{SO}}$. The latter originates from the orbital accumulation, hence we denote these as spin and orbital contributions to the magnetic torque respectively.
%
The magnetic torque in 2D systems is often described by non-equilibrium spin accumulation, but we emphasize that its origin is the non-equilibrium orbital angular momentum generation.

We assess the two contributions to the torque using a first-principles description of the system's electronic structure, and Kubo formalism for electic field response $\mathbf{E}$, providing the details of our calculations in Supplemental Material~\cite{Supplementary}. We use the code FLEUR \cite{fleur}, which implements full-potential linearized augmented plane wave method \cite{Wimmer1981}, in combination with Wannier interpolation \cite{Freimuth2008, Pizzi2020} to efficiently compute the anti-damping part of the torkance tensor $\hat{T}$ defined as $\boldsymbol{\mathcal{T}}=\hat{T}\,\mathbf{E}$. When represented in spherical coordinates we are interested only in the polar $\theta$-component of the torkance tensor, where $\theta$ is the angle that the magnetization makes with the $z$-axis, see Fig.~\ref{fig:OutofPlaneInPlane}.
 Correspondingly, we assume simplified notations of $T^{\theta}_\mathbf{M}$, $\varPhi^\theta$ and $T_{\rm SO}^\theta$ for the $\theta$-component of the torkance tensor, and its spin-flux and orbital parts. When studying the angular dependence of the torque, we consider only the case when the magnetization is tilted away from the $z$-axis into the $x$-axis while being kept in the $xz$-plane.
We apply the external electric field along $x$.

We first compute the magnitude of the total torkance and its decomposition in a single A-layer of FGT as a function of angle $\theta$, shown in  Fig.~\ref{fig:AngularDepSOT}(a). As predicted by symmetry~\cite{Johansen2019}, single-layer FGT hosts a non-vanishing SOT for $\theta\neq 0$. Interestingly, we find a sign change at $\theta \approx 60^{\circ}$ with the largest torkance when $\mathbf{M}$ is in-plane, Fig.~\ref{fig:AngularDepSOT}(a). Markedly, the single-layer SOT is entirely orbital in origin as we find a complete lack of flux torkance for all magnetization angles. 
For the A-layer in bulk, Fig.~\ref{fig:AngularDepSOT}(b), the angular dependence has a similar functional form to single layer FGT, with an overall suppression in magnitude. 
What stands in stark contrast to single-layer FGT is that the spin contribution,
$\varPhi^{\theta}$, constitutes a major component to the torque. This suggests that the bulk spin contribution may originate from the adjacent single-layer orbital contribution.
However, as $\theta$ gets closer to $90^\circ$, the spin contribution $\varPhi^\theta$ is strongly suppressed and the orbital contribution $T_\mathrm{SO}^\theta$ becomes dominant. This results in a peculiar angular dependence of the magnetic torque, which is significantly different from the prediction based on the leading order symmetry expansion~\cite{Johansen2019, Kurebayashi2022}. We also remark that bulk FGT is the first example of a material exhibiting orbital-spin crossover of the magnetic torque as a function of magnetization angle. This can be attributed to the 2D nature of the crystal structure, which results in highly anisotropic electronic structure. 


In order to understand the origins of spin and orbital contributions to the magnetic torque, we calculate non-equilibrium spin and orbital accumulation induced by an external electric field, shown in Fig.~\ref{fig:AngularDepSOT}(c) for the $y$-component which is relevant for $T_\mathbf{M}^\theta$. For small values of $\theta$, both $ S_{y} $ and $L_{y}$ are proportional to $M_x$, which is consistent with symmetry analysis in the lowest order of $\mathbf{M}$ \cite{Johansen2019}. However, as $\theta$ increases, higher order contributions become more pronounced. Interestingly, $S_{y}$ and $L_{y}$ exhibit a qualitatively different angular dependence: $S_{y}$ changes sign when $\theta$ is between $30^\circ$ and $45^\circ$, while $L_{y}$ monotonically increases as $\theta$ increases. Moreover, $L_{y}$ is an order of magnitude larger than $ S_{y} $ over a wide range of $\theta$. This strongly supports the idea that orbital accumulation significantly contributes to the magnetic torque, especially for large values of $\theta$. To understand why for smaller angles the orbital accumulation does not translate into a large orbital torque, we proceed to investigate the state-resolved properties to shed light on the microscopics of the crossover behaviour.


We analyze band-resolved contributions to $T^{\theta}_{\mathbf{M}}$ for $\theta=30^\circ$ [Fig.~\ref{fig:TorquesOnBandsAndEDepSOT}(a)] and $90^\circ$ [Fig.~\ref{fig:TorquesOnBandsAndEDepSOT}(b)]. These two angles are chosen such that they exhibit the strongest $\varPhi^{\theta} $ and $T_\mathrm{SO}^{{\theta}}$, respectively. As expected, the distributions of $T^{\theta}_\mathbf{M}$ for $\theta=30^\circ$ and $\theta=90^\circ$ are qualitatively different. In particular, we find that the torque is spread over wider regions of phase-space for $\theta=30^\circ$, while for $\theta=90^\circ$ the torque originates from isolated contributions, especially for energies above the Fermi level. To characterize the role of orbital accumulation we plot the state-resolved $L_{y,n\mathbf{k}}$ at $\theta=30^\circ$ [Fig.~\ref{fig:TorquesOnBandsAndEDepSOT}(c)] and $\theta=90^\circ$ [Fig.~\ref{fig:TorquesOnBandsAndEDepSOT}(d)]. By comparing Figs. \ref{fig:TorquesOnBandsAndEDepSOT}(a) and \ref{fig:TorquesOnBandsAndEDepSOT}(c), we immediately notice that the isolated hotspot-like regions for $T^{\theta}_\mathbf{M}$ and $ L_y $ at $\theta=30^\circ$ are either not in the same position or have a different sign, which is especially visible around the Fermi energy. This is expected from Fig. \ref{fig:AngularDepSOT}(b) because the magnetic torque is dominated by the spin flux at $\theta=30^\circ$. On the other hand, at $\theta=90^\circ$ the correlation between $T^{\theta}_\mathbf{M}$ and $L_y$ is much more pronounced [compare Figs. \ref{fig:TorquesOnBandsAndEDepSOT}(b) and \ref{fig:TorquesOnBandsAndEDepSOT}(d)]. 
 
To understand the suppression of orbital torque despite prominent current-induced orbital magnetization of states at small angles, we compare the $\mathbf{k}$-resolved contribution for $L_y$ summed over all occupied states at $\theta=30^\circ$ [Fig.~\ref{fig:TorquesOnBandsAndEDepSOT}(g)] and $\theta=90^\circ$ [Fig.~\ref{fig:TorquesOnBandsAndEDepSOT}(h)]. While the summation over $\mathbf{k}$ leads to an overall orbital moment at $\theta=30^\circ$, the orbital moment at $\theta=90^\circ$ is evidently greater, fitting with Fig.~\ref{fig:AngularDepSOT}(c). Furthermore, there is asymmetry present at $\theta=30^\circ$ that is not at $\theta=90^\circ$, implying a crossover of orbital characters that could drive the orbital torque. The corresponding asymmetry of the orbital distribution for $\theta=30^\circ$ can be also directly seen in the band structure plot for energies above the Fermi level, Fig. \ref{fig:TorquesOnBandsAndEDepSOT}(c). Conversely, the corresponding current induced spin distribution, presented in Supplemental Material~\cite{Supplementary}, displays the opposite behavior. These findings help us understand the intricate energy-dependent interplay between the spin and orbital components to the magnetic torque as a function of the tilting angle. 

\textcolor{black}{In previous works, the effect of strong correlations using dynamical mean field theory are investigated \cite{Kim2018b}. Within the energy range chosen in our investigation, the only key difference between both methods is the position of the Fermi level.} Hence, we show the Fermi energy dependence of $T^{\theta}_\mathbf{M}$, $\varPhi^{\theta}$, and $ T_\mathrm{SO}^{{\theta}}$ at $\theta=30^\circ$ and $90^\circ$ in Figs. \ref{fig:TorquesOnBandsAndEDepSOT}(e) and \ref{fig:TorquesOnBandsAndEDepSOT}(f), respectively. At $\theta=30^\circ$, $\varPhi^{\theta}$ is the main contribution to $T^{\theta}_\mathbf{M}$ although $T_\mathrm{SO}^{{\theta}}$ is also non-vanishing [Fig. \ref{fig:TorquesOnBandsAndEDepSOT}(e)]. On the other hand, at $\theta=90^\circ$, $T^{\theta}_\mathbf{M}$ is mainly governed by $ T_\mathrm{SO}^{{\theta}}$ and $\varPhi^{\theta}$ is suppressed over a wide energy range [Fig. \ref{fig:TorquesOnBandsAndEDepSOT}(f)]. In Figs. \ref{fig:TorquesOnBandsAndEDepSOT}(a-d), we notice that the hotspots are mainly concentrated around the nodal lines near $\Gamma$ at $+0.2\ \mathrm{eV}$ \cite{Kim2018b}. This implies that the torkance can be efficiently tuned with band filling. Indeed, we find a substantial increase of $T^{\theta}_\mathbf{M}$ as the Fermi energy is raised by $\sim 0.2\ \mathrm{eV}$ at both $\theta=30^\circ$ and $90^\circ$. Interestingly, at $\theta=30^\circ$, $T^{\theta}_\mathbf{M}$ changes sign from negative to positive as the Fermi energy is increased by $\approx 0.2\ \mathrm{eV}$. On the other hand, at $\theta=90^\circ$, the sign of $T^{\theta}_\mathbf{M}$ remains positive. This suggests a possibility of tuning the sign and magnitude of the torkance by doping. Such complex anisotropic torkances can be harnessed to drive the non-trivial excitations of spin textures that FGT hosts~\cite{Hanke2020}. According to our estimate, raising the energy by $0.2\ \mathrm{eV}$ requires $\approx1.5$ additional electrons per layer. This may be achieved e.g.~by substitutional doping of Fe by Co. 

Let us comment lastly on the relevance of our predictions for the spin-orbit torque measurements on realistic FGT samples. While the single-layer SOT in bulk samples can be promoted if global symmetry is broken~e.g.~by surfaces and interfaces~\cite{Martin2021, Zhang2021}, 
inversion symmetry present in ideal bulk FGT ultimately suppresses the overall torque on the magnetization. However, the reality is that there are a variety of ``intrinsic'' mechanisms which can drive the inversion symmetry breaking in this bulk system~\cite{Narita2018,Martin2021,Chakraborty2022}. 
One of these effects is that of magnetostriction \cite{Narita2018}. Here, due to the coupling of the lattice to the magnetic moment, the rotation of the moment can initiate a crystal structure change, breaking the inversion symmetry of the perfect crystal. \textcolor{black}{Similarly, the effects of the electric field on the system could give rise to high frequency magnon modes capable of canting the spins and driving a torque even in the pristine limit}. It is also plausible to assume that due to the 2D vdW nature of the material it is possible for the layers to slip over one another. In some instances, if multiple pairs of layers have displaced in different ways, the inversion symmetry could be broken in this fashion. 
Moreover, recent experimental work reports a discrepancy in the occupancy of the central Fe atom in the A layer versus the B layer~\cite{Chakraborty2022}. This implies that the unit cell is no longer inversion symmetric providing an avenue for bulk torque generation. Our results can be in principle readily applied to this situation:  Due to the inter-layer occupancy difference being small, from the theoretical perspective the corresponding torque should be simply given by the difference in the torques of layers A and B from Fig.~\ref{fig:TorquesOnBandsAndEDepSOT}(e,f) for corresponding occupancies, and the corresponding effective spin-orbit field can be found in a similar fashion~\cite{Johansen2019}. 

\textcolor{black}{Finally, the determination of the components of the spin and orbital torque remain a significant challenge. For example, standard techniques for measuring the current-induced torques only measure the total torkance, but this can be overcome by attempting to find an `even'-`odd' effect such as in MoS$_2$ \cite{Cysne2021} by consecutively removing vdW layers. 
To probe the orbital accumulation, a key signature of orbital-driven torkances, one can use optical probes such as magneto-optical Kerr effect~\cite{Choi2021} or x-ray magnetic circular dichroism~\cite{Stamm2019}. }

To summarize, in our work, we uncovered that in magnetic vdW materials the microscopics of spin-orbit torques can be extremely rich owing to the pronounced 2D nature of constituting layers and their orbital complexity. As we show for FGT, this ultimately results in profoundly anisotropic torque properties, which may be mediated by an exotic  orbital-spin crossover. Given the fundamentally different properties of spin and orbital angular momentum out of equilibrium, this suggests exciting possibilities in exploring spin-orbital dynamics of 2D magnets and corresponding transport manifestations. We further argue that the detailed knowledge of ``hidden" spin-orbit torques in bulk vdW materials can provide a key to designing their current response via educated symmetry breaking, which may prove pivotal for the integration of 2D magnets into the spintronic device setting.


\begin{acknowledgments}
The authors appreciate fruitful discussions with Prof. Arne Brataas, Dr. Martin Gradhand, Maximilian Merte, Fabian Lux and Dr. Sumit Ghosh. We gratefully acknowledge the J\"ulich Supercomputing Centre for providing computational resources under project jiff40. This work was funded by the Deutsche Forschungsgemeinschaft (DFG, German Research Foundation) $-$ Grants No. TRR 173/2 $-$ 268565370 Spin+X (projects A01, BO2 and A11), No. TRR 288 $-$ 422213477 (project B06) and CRC 1238 -- 277146847 (project C01).
\end{acknowledgments}

\end{document}